# Dynamically Tunable Membrane Metasurfaces for Infrared Spectroscopy

Furkan Kuruoglu[1,2+], Samir Rosas[1+], Jin-Woo Cho[3], David A. Czaplewski[4], Yuri Kivshar[5], Mikhail Kats[3], Filiz Yesilkoy[1]*

[1]Department of Biomedical Engineering, University of Wisconsin-Madison, Madison, Wisconsin 53706, USA
[2]Department of Physics, Faculty of Science, Istanbul University, Vezneciler, 34134, Istanbul, Turkey
[3]Department of Electrical and Computer Engineering, University of Wisconsin-Madison, Madison, Wisconsin 53706, USA
[4]Center for Nanoscale Materials, Argonne National Laboratory, Lemont, Illinois 60439, USA
[5]Nonlinear Physics Center, Research School of Physics, Australian National University, Canberra ACT 2601, Australia
Corresponding Author e-mail: filiz.yesilkoy@wisc.edu
[+]equal contribution

## Abstract

Mid-infrared spectroscopy enables biochemical sensing by identifying vibrational molecular fingerprints, but it faces limitations in instrumentation portability and analytical sensitivity. Optical metasurfaces with strong mid-IR photonic resonances provide an attractive solution towards on-chip spectrometry and sensitive molecular detection, yet their static nature hinders their anticipated impact. Here, we introduce and demonstrate dynamically tunable silicon membrane metasurfaces exhibiting high-Q transmissive resonances in the fingerprint region. By harnessing silicon's thermo-optical properties, we achieve continuous modulation of electromagnetically induced transparency (EIT)-like modes that emerge upon the interference of quasi-bound states in the continuum (q-BICs) and surface lattice modes. We measure a spectral tuning rate of 0.06 cm$^{-1}$/K by continuously sweeping the sharp EIT resonances over a 23.5 cm$^{-1}$ spectral range across a temperature range of 300-700 K. This dynamic transmission control enables non-contact chemical analysis of polymer films by detecting characteristic absorption bands of polystyrene (1450 and 1492 cm$^{-1}$) and Poly(methyl methacrylate) (1730 cm$^{-1}$) without bulky spectrometers. When analyte molecules fill the metasurface-generated photonic cavities, we demonstrate vibrational strong coupling between the Poly(methyl methacrylate)'s carbonyl band and the EIT mode, manifested in the Rabi splitting of ~43 cm$^{-1}$. Our results establish a new photonic platform that unites spectral precision, strong field enhancement, and reconfigurability, offering diverse potential for compact mid-IR spectroscopy, molecular sensing, and programmable polaritonic photonics.

**Keywords:**

tunable metasurface, thermally reconfigurable metasurface, mid-infrared photonics, bound states in the continuum (BICs), electromagnetically induced transparency (EIT), surface-enhanced infrared spectroscopy (SEIRAS), vibrational strong coupling, chemical sensing

# Introduction

The mid-infrared (mid-IR) spectral range cultivates vibrant photonics research driven primarily by the pressing needs of biochemical sensing and molecular spectroscopy applications. Absorption spectroscopy enables rapid and precise chemical identification by detecting molecules' distinctive vibrational signatures in the mid-IR region, commonly referred to as fingerprint spectra. Research and development in industrial and academic settings relies heavily on mid-IR spectroscopy using Fourier transform infrared (FTIR) spectrometers and the more recent laser-based discrete frequency interrogation systems. Despite their analytical strength within controlled laboratory settings, these instruments exhibit fundamental constraints, such as large footprints, operational complexity, and high acquisition costs, that obstruct their deployment for widespread, field-based sensing applications. Furthermore, conventional IR spectroscopic techniques typically demonstrate insufficient sensitivity for trace analyte detection due to the inherently limited absorption cross-sections of molecular vibrations ($10^{-20} \sim 10^{-16}$ cm$^2$)[1]. These technological and analytical challenges call for innovative photonic approaches to develop field-deployable, highly sensitive mid-IR analytical platforms capable of addressing urgent practical challenges across biomedicine, environmental monitoring, and advanced manufacturing fields.

Optical metasurfaces with engineered subwavelength architectures enable versatile manipulation of IR light, offering transformative solutions for mid-IR absorption spectroscopy[2,3]. To develop compact and spectrometer-less instrumentation, plasmonic mid-IR metasurfaces have been employed as filter arrays to endow microbolometer cameras with spectral selectivity[4–6]. This approach leverages intensity variations in camera pixel readouts to reconstruct spectra or to inform ML-based chemical sensing without bulky instrumentation. Other techniques have combined plasmonic perfect absorbers with pyroelectric materials for wavelength-selective IR detectors optimized for compact gas sensors[7]. Beyond their far-field narrow-band filter implementations, metasurfaces have advanced biochemical sensing through their near-field light localization capabilities into subwavelength volumes called hotspots. To date, surface-enhanced infrared spectroscopy (SEIRAS) has enabled important biomedical applications, including living cell[8], tissue[9], lipid membrane[10], and human fluid analysis[11]. Moreover, high quality factor (Q) resonances in dielectric metasurfaces, such as arrays of low-loss Si/Ge structures[12,13] and perforated free-standing Si membranes[14,15], significantly enhance light-matter interactions across weak to strong coupling regimes[16], thereby improving molecular detection[14,15]. However, the static nature of these metasurfaces presents a significant limitation that their high-Q resonances restrict cavity-coupled enhancement to narrow spectral ranges at specific spatial positions, highlighting the need for reconfigurable mid-IR metasurfaces.

Dynamically tunable metasurfaces are highly sought-after components in photonics, because they enable real-time control over resonance frequencies, spectral bandwidths, and wavefront shaping[17–20]. Previously, electrically tunable mid-IR metasurfaces were proposed using GaAs[21,22], graphene[23–27] and other van der Waals materials[28,29] for SEIRAS applications. However, a combination of challenges associated with large-area fabrication of two-dimensional materials and broad resonances created by surface plasmon and phonon polaritons hindered the practical implementations of these devices for chemical sensing. Alternatively, phase-change materials (PCMs), such as Ge$_2$Sb$_2$Te$_5$ (GST)[30–32], In$_3$SbTe$_2$[33] and VO$_2$[34], have been leveraged to endow plasmonic and dielectric metasurfaces with dynamic resonance tunability in the mid-IR. Yet this approach requires incorporating PCMs within the near-field hotspots of resonators to utilize refractive index modulation for active tuning—a configuration fundamentally incompatible with conventional metasurface sensing methodologies. Furthermore, the inherent lossy characteristics of PCMs in the mid-IR region compromise the high-Q resonances of the dielectric resonators. These limitations underscore the critical need for novel photonic approaches to develop reconfigurable high-Q mid-IR metasurfaces optimized for chemical sensing and spectroscopic applications.

Here, we introduce dynamically tunable free-standing Si membrane metasurfaces exhibiting high-Q transmissive resonances in the mid-IR region, functioning as on-demand spectrally selective filter arrays, with significant implications for advanced mid-IR spectroscopy. By harnessing Si's thermo-optical properties, i.e., its refractive index changes with varying temperatures, we achieve continuous modulation of high-Q (maximum measured 187 @ 1240 cm$^{-1}$) transmission modes in the fingerprint spectral range at a tuning rate of 0.06 cm$^{-1}$/K. Our metasurface design supports an analog of electromagnetically induced transparency (EIT) resonance, arising from the interference of broad surface lattice modes (SLMs) and ultra-sharp quasi-bound states in the continuum (q-BIC) resonances. Through the spatiotemporal modulation of these resonances, we present two sets of experimental demonstrations: i) the spectrometer-free, high-resolution spectral fingerprinting of thin polymer films, Poly(methyl methacrylate) (PMMA) and polystyrene (PS), remotely positioned from the metasurface, ii) vibrational strong coupling between PMMA molecules and dynamically swept metasurface resonances as they traverse the molecule's fundamental vibrational mode. Our results show that real-time tunability of sharp mid-IR resonances enables chemical analysis of remote objects. Moreover, ultra efficient near-field light localization within accessible air voids leads to quantum coherent light-matter interactions, with profound implications for ultrasensitive molecular detection. With promising advancements towards compact analytical instrumentation and enhanced sensing capabilities of the mid-IR spectroscopy, our approach has the potential to address the critical analytical needs of biomedicine, industrial manufacturing, and environmental safety.

**Results**

To investigate the thermal tunability of the Si membrane metasurfaces, we performed temperature-dependent optical characterization using a broadband mid-IR source integrated with the FTIR microscope and an electrically controlled heating stage, as illustrated in Figure 1a. Our photonic device supports a metasurface analog of electromagnetically induced transparency (EIT) resonance, wherein a broad multipolar surface lattice mode (SLM) interferes with a narrow quasi-bound state in the continuum (q-BIC) mode. The q-BIC mode emerges through in-plane symmetry breaking, achieved by oppositely tilting an aperture rod pair in a meta-unit, as shown in Figure 1c. When these two interfering resonances spectrally and spatially overlap, a narrow transparency window emerges in the mid-IR spectrum, which we designate as q-BIC-EIT mode[15] . The arrayed tilted rods were patterned through the 1 μm thick single-crystalline Si membrane by electron beam lithography and anisotropic dry etching, as detailed in the Methods section (Figure 1b). The intrinsic thermo-optical properties of crystalline Si facilitate controlled modulation of its complex refractive index[35,36] upon thermal stimulation, introducing a pronounced spectral shift of the q-BIC-EIT resonance, as depicted in Figure 1d. By spectrally sweeping the transmissive resonance peaks across the fingerprint spectra, we detected the characteristic absorption bands of analytes, whether positioned remotely from the metasurface or directly interfaced for enhanced near-field interactions.

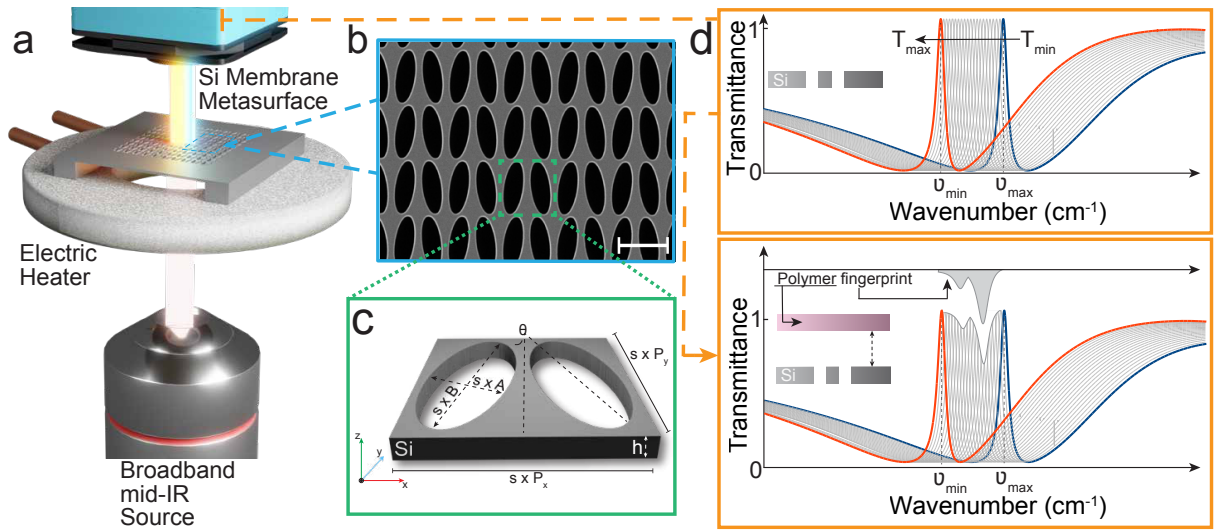

**Figure 1: Dynamically tunable Si membrane metasurfaces for mid-IR spectroscopy. a)** Schematic of the optical characterization system in transmission mode, where a mid-infrared source illuminates the metasurface placed on an electrically controlled heating stage and transmission spectra are collected as the temperature of the stage is varied, **b)** Top-view SEM image of the fabricated Si membrane metasurface, illustrating the periodic perforations that form its lattice (scale bar= 5 μm), **c)** Three-dimensional schematic of the meta-unit, highlighting the key geometric parameters, including the periodicities ($P_x$, $P_y$), membrane thickness ($h$), elliptical tilted rod apertures with major and minor axes ($A$, $B$), tilting angle (Theta) and geometric scaling factor ($s$), **d)** Temperature-dependent transmission spectra of the metasurface as a function of wavenumber ($v$). A continuous redshift in the resonance position is observed at a tuning rate of 0.06 cm$^{-1}$/K, due to the temperature-dependent refractive index variation of the Si membranes. The dynamic resonance tunability enables to capture of characteristic fingerprints of the analytes, whether positioned remotely from the metasurface or directly interfaced for enhanced near-field interactions.

To study the dynamic tunability of q-BIC-EIT resonances with varying temperatures, we performed finite element method simulations using COMSOL Multiphysics software. The calculated transmittance spectra were plotted in Figure 2a for a metasurface with $\theta$=8° rod tilting angle by varying the temperature from 300 K (room temperature, RT) to 700 K with 50 K steps. A spectral shift of 25.6 cm$^{-1}$, greater than the full width at half maximum (FWHM) of the resonance (11.4 cm$^{-1}$), was achieved with a temperature difference of 400 K. In the elevated temperature range, the transmittance of the membrane metasurface exhibited a reduction (9 % at 700 K compared to RT) due to the increased material losses. To elucidate the impact of temperature on the q-BIC-EIT resonance characteristics of the membrane metasurfaces, we plotted the electric (E) field enhancement and displacement current profiles for the simulated minimum (300 K) and maximum (700 K) temperatures at the resonance. Figure 2b shows that the E-field enhancement ($E_{enh}$=|E|/|$E_0$|) maintains a consistent profile at varying temperatures, with hotspots emerging at the tips of the elliptical apertures. To further quantify the temperature dependent E-field enhancement, we plotted the maximum $E_{enh}$ in a metaunit as a function of temperature for different rod tilting angles ($\theta$=2°, 4°, 6°, and 8°) in Figure 2c. The $E_{enh}$ is higher for smaller rod tilting angles across the entire temperature range, a characteristic of q-BIC modes supporting higher Q-factor resonances at low asymmetry conditions. At high temperatures (T>500 K), $E_{enh}$ decreases consistently for all $\theta$ values, with a higher gradient for smaller $\theta$. We calculated 25.8, 5.7, 1.3 and 0 % $E_{enh}$ decrease at 700 K compared to 300 K, for the $\theta$=2°, 4°, 6°, and 8°, respectively.

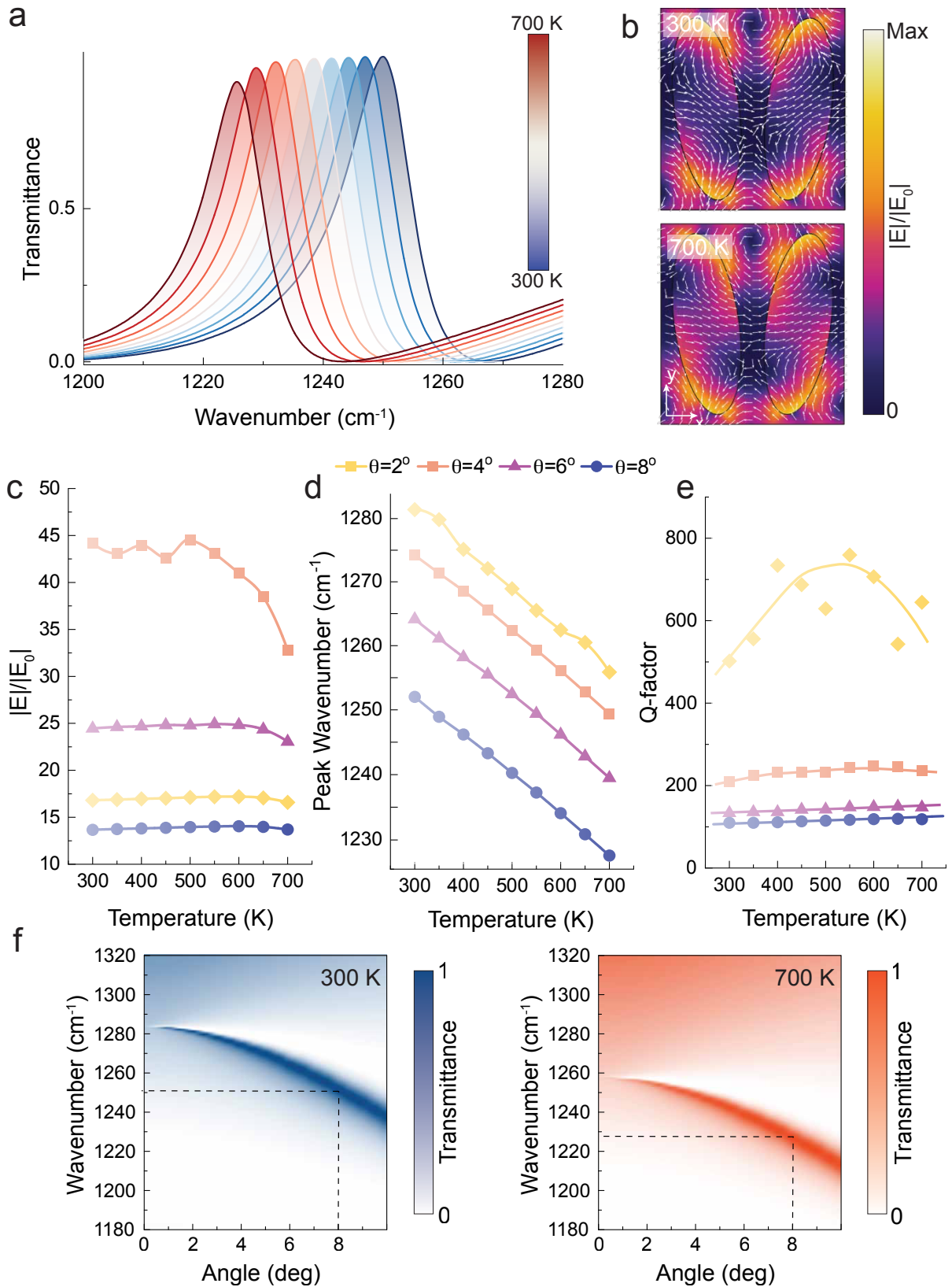

**Figure 2: Thermal tuning on hybrid resonant modes. a)** Numerically simulated transmittance spectra of the Si membrane metasurface with $\theta = 8°$ tilted elliptical rods supporting photonic analog of EIT resonances. Spectral resonance tunning was achieved by varying the metasurface temperature from 300 K to 700 K. **b)** The electric (E)-field enhancement maps and displacement current density arrow profiles on a meta-unit surface show that the E-field enhancement profile remains the same for different temperatures. **c)** The maximum E-field enhancement as a function of temperature. $|E|/|E_0|$ degrades at higher temperatures for all tilting rod angles, with a higher gradient at smaller tilting angles. **d)** The resonance peak wavenumber shifts as a function of temperature, showing thermal

tuning rates for each rod tilting angle. **e)** The Q-factor of the resonance as a function of temperature. **f)** The transmittance colormaps show the overall effects of the temperature on the spectral shift and of the q-BIC-EIT modes supported by the Si membrane metasurfaces.

Figure 2d and 2e further depict the temperature-dependent resonance characteristics, showing both the peak wavenumber and the Q-factor variations as functions of temperature. As the temperature increases, a clear redshift in the resonance position is observed for each tilting angle. This trend indicates the underlying thermo-optical effect in Si, where the temperature-dependent change in the real and imaginary parts of the refractive index modulates the spectral position and linewidth of the resonance. The slope of the resonance shift reveals the highest resonance tunability as 0.064 cm$^{-1}$/K with a rod tilting angle of $\theta=2°$. Moreover, the tunability rate decreases by 5.61 % for $\theta=8°$ metasurface compared to $\theta=2°$. This result is consistent with the known characteristics of the q-BIC resonances where the increased rod tilting angles introduce additional radiative losses, thereby reducing the sensitivity of the resonance to temperature-induced refractive index variations. Colormaps of transmittance as a function of wavenumber and rod tilting angle at 300 K and 700 K are presented in Figure 2f, providing a comprehensive visualization of the temperature and geometrical-asymmetry dependent q-BIC-EIT resonance evolution in our Si membrane metasurfaces.

We fabricated Si membrane metasurfaces with varying rod tilting angles and measured their temperature-dependent resonance characteristics to compare with the simulation predictions. Figure 3a shows the mid-IR transmittance spectra obtained from fabricated metasurfaces with rod tilting angles $\theta=6°$ and $0°$, corresponding to q-BIC-EIT and SLM, where the q-BIC vanishes to BIC and the transparency window disappears. While the temperature-dependent spectral shifts are evident in both modes, the q-BIC-EIT mode is more sensitive to temperature modulation than SLM. Figure 3b illustrates that all measured metasurfaces with varying rod tilting angles exhibit a consistent linear shift in their resonances across the examined temperature range (300-700 K). In strong agreement with the simulation results, the high-Q metasurface ($\theta=2°$) demonstrates the highest tunability rate at 0.06 cm$^{-1}$/K, while the tunability rate of the SLM ($\theta=0°$) is measured as 0.051 cm$^{-1}$/K. Figure 3c shows the experimentally derived Q-factors from metasurfaces with varying rod tilting angles at the lowest (300 K) and highest (700 K) measured temperatures. The inverse quadratic relationship of the Q factor with the asymmetry parameter ($\theta$) indicates the q-BIC nature of the resonances and shows consistency independent of the temperature variations[37].

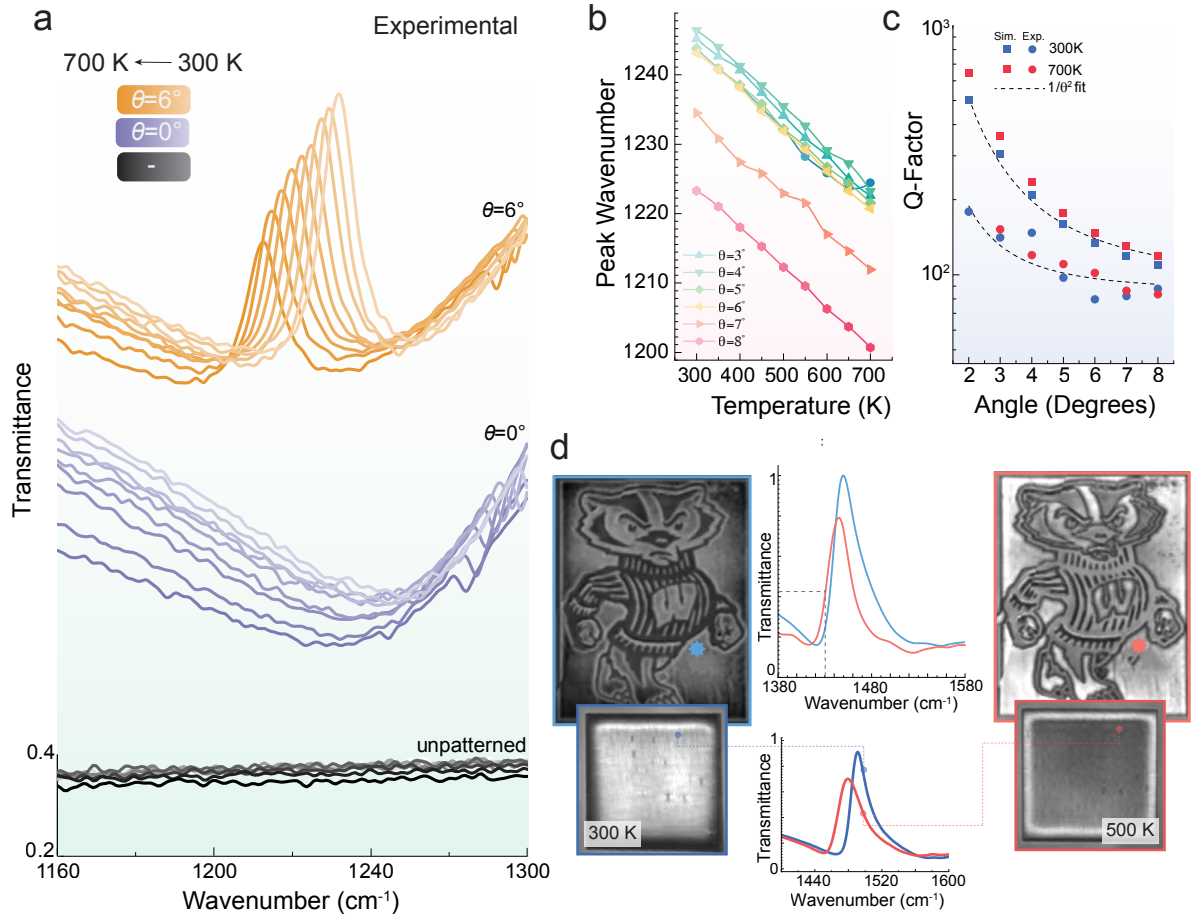

**Figure 3: Experimentally measured thermal tuning effects on the photonic q-BIC-EIT modes. a)** Temperature tunability characteristics of the Si membrane metasurfaces with $\theta = 6°$ (yellow lines) showing q-BIC-EIT mode and $\theta = 0°$ (violet lines) showing SLM where q-BIC vanishes to BIC, both red shifting as the real part of the refractive index increases with increasing temperature. Black lines show transmission through unpatterned membrane. **b)** Measured resonance peak wavenumber as a function of temperature for different tilted rod angles. For all angles, the measured thermal tuning rate is $\Delta v/\Delta T \approx -0.05$ cm$^{-1}$/K. **c)** Evolution of Q-factor as a function of rod tilting angle $\theta$ for simulation and experimental data, showing inverse quadratic correlation (Q-factor $\propto \theta^{-2}$) for different temperatures. **d)** Wide-field IR microscope images of the UW-Madison mascot, Bucky Badger, patterned metasurface, measured by illuminating at 1430 cm$^{-1}$ at 300 K and 500 K temperatures. As the temperature increases, the resonance peak shifts to lower wavenumbers, leading to a contrast inversion as also indicated on the FTIR measured transmittance spectra. Similar contrast inversion effect is shown on a uniform square metasurface, where the bright to dark transition occurs at 1498 cm$^{-1}$ as temperature increases.

To visualize the dynamic resonance tunability characteristics, in Figure 3d, we show the mid-IR microscope images of a partially patterned metasurface in the shape of Bucky Badger, UW-Madison's mascot. The Bucky metasurface images were captured at 1430 cm$^{-1}$ illumination in transmission mode at two different temperatures (300 K and 500 K). Due to a dynamic redshift in the resonance peak upon temperature increase, metasurface patterned regions become more transmissive at the same illumination wavenumber (1430 cm$^{-1}$) revealing a dark-to-bright contrast transition. This effect is also depicted in the transmittance spectra measured at the indicated red and blue star positions of the images at 300 K and 500 K. With only a 200 K temperature increase, a notable transmittance boost from 0.23 to 0.39 (68.2%) was measured. We also show an opposite bright-to-dark transition by probing the resonance peak at 1498 cm$^{-1}$, illustrated in the lower segment of Figure 3d. Here, a uniformly patterned Si membrane area of 400 µm x 400 µm exhibits a 58.7 % decrease in transmittance as a result of a 200 K temperature increase.

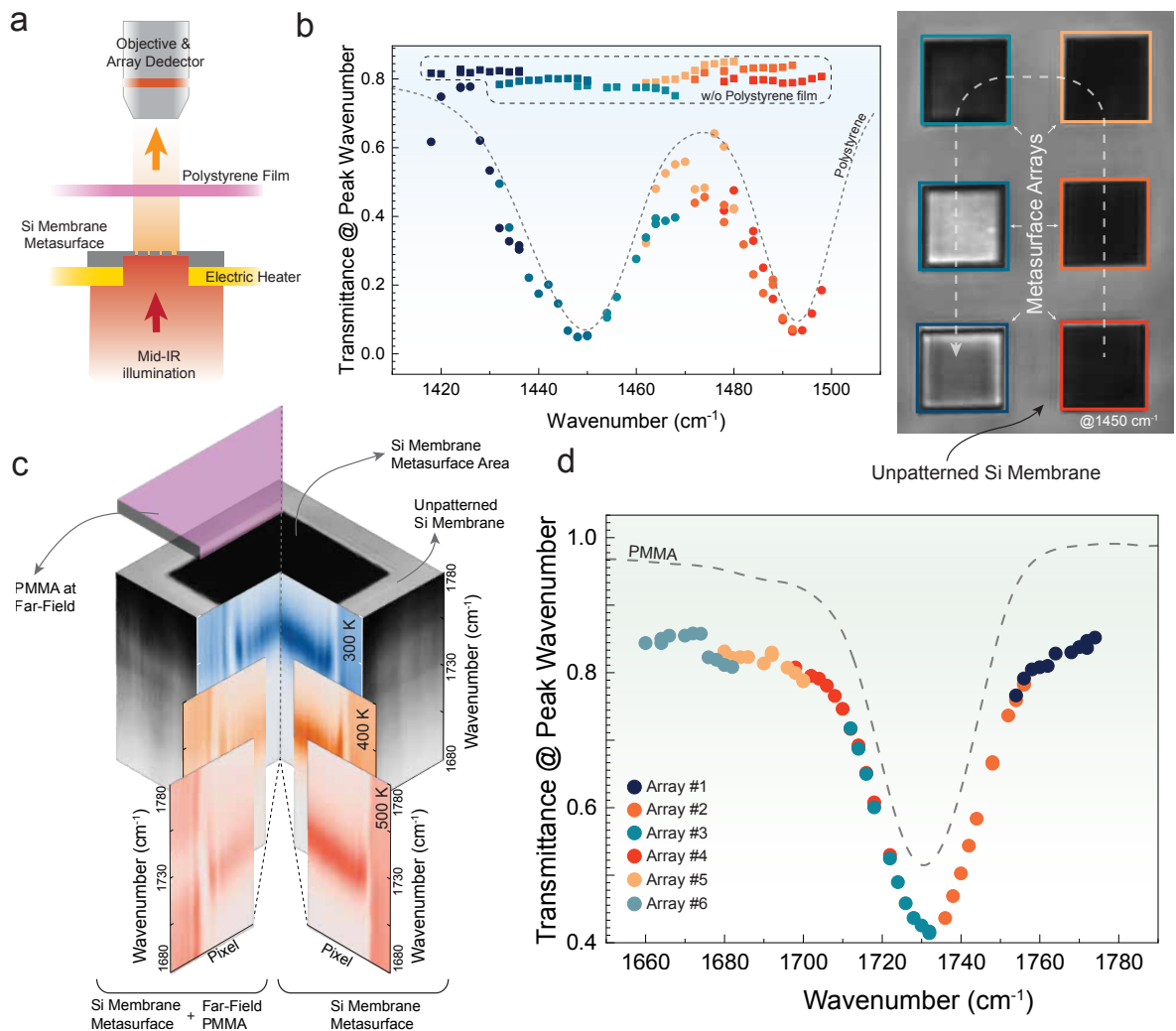

**Figure 4: Chemical analysis of remotely positioned polymer films using dynamically tunable metasurfaces.**
**a)** Schematic illustration of the material analysis setup where a polymer film is placed at least 10 mm distant from the thermally tuned metasurface. **b)** The maximum transmittance of six different metasurfaces were measured at varying temperatures from 300 to 500 K, with and without a polystyrene film (thickness=38 μm). Each metasurface covers around 20 cm$^{-1}$ spectral range upon 200 K temperature sweep. Six metasurfaces were sufficient to capture absorption fingerprints of the polystyrene including modes at 1450 and 1492 cm$^{-1}$ corresponding to C=C stretching vibrations. The colors of the data points correspond to the colored frames of the metasurfaces in the IR microscope image captured at 1450 cm$^{-1}$. This data shows how six discrete metasurfaces can capture a continuous fingerprint spectrum enabled by the dynamic resonance tuning capability. **c)** The hyperspectral data visualization of a single metasurface showing temperature-dependent spectral shifts at three discrete temperatures, 300 K, 400 K, and 500 K. The left panel shows measured data with PMMA film, whereas the right panel depicts the bare metasurface. Spectral colormaps effectively demonstrate spectral homogeneity across the metasurface area, both in the presence and absence of the PMMA film. Corresponding to the PMMA C=O absorption band at 1730 cm$^{-1}$, the resonance transmission of the metasurface decreases as it is thermally swept through the band. **d)** The maximum transmittance of six metasurfaces each with a different resonance wavenumber are measured at varying temperatures from 300 to 500 K, with and without a PMMA film of 700 nm thickness.

We leveraged the temperature tunability of the Si membrane metasurfaces as dynamic narrow-band mid-IR filters to demonstrate non-contact chemical analysis of polymer films. Figure 4a depicts the transmission-mode optical imaging configuration, wherein each metasurface's spectral response was characterized while modulating the substrate temperature. An example of this chemical analysis is shown in Figure 4b, where we employed an array of six metasurfaces with predetermined resonance wavenumbers assigned during fabrication. When a polymer film specimen is inserted into the optical path, the characteristic vibrational absorption signatures of the molecular constituents induce distinct transmission intensity variations across the metasurface array elements, each exhibiting different magnitudes of attenuation as their resonance frequencies are thermally tuned. The frame color of each metasurface in the mid-IR image corresponds to the color of the spectral data points presented in the accompanying plot. Square and circular data points represent the transmittance values of each metasurface at varying temperatures in the absence and presence of a PS film, respectively. By continuously sweeping the metasurface array temperature from 300 to 500K, we acquired a continuous fingerprint spectrum of PS, clearly resolving the 1450 cm$^{-1}$ and 1492 cm$^{-1}$ absorption bands associated with (C=C stretching) benzene ring vibrations, respectively. Similarly, we demonstrated detection of a thin PMMA film, successfully resolving its characteristic carbonyl (C=O) stretching vibration at 1730 cm$^{-1}$ (Figure 4d). For both polymer species, our approach accurately determined the spectral position and FWHM of the absorption bands, with measured transmission modulation aligning closely with reference spectra obtained through conventional spectroscopy.

To characterize the spatial characteristics of our thermally tunable metasurfaces, we used a hyperspectral imaging method. Figure 4c presents the spectral images acquired with and without a PMMA film at discrete temperatures of 300, 400 and 500 K. Across the metasurface patterned membrane region (dimensions 400 x 400 $\mu m^2$), the transmittance magnitude and resonance wavenumber exhibited spatial uniformity, indicating a homogeneous fabrication process and even thermal distribution. Spatially resolved transmittance profiles within the metasurface element show the shift in resonance peak toward 1730 cm$^{-1}$ upon temperature increase from 300 to 500 K. The spectral overlap of the metasurface resonance and the characteristic carbonyl absorption band of PMMA induces a uniform transmission attenuation across the entire metasurface. Notably, imaging-based interrogation holds significant potential for chemical investigation of spatially heterogeneous specimens through pixel-by-pixel analysis of the focal plane array detector data.

To investigate the near-field light-matter interactions on the thermally modulated metasurfaces, we deposited PMMA thin films onto Si membrane metasurfaces. Leveraging the low material dissipation and high-Q resonances of the Si membrane metasurfaces, we previously demonstrated vibrational strong coupling (VSC) between the PMMA's carbonyl (C=O) band and q-BIC and q-BIC-EIT resonance modes. Here, we study coherent energy exchange dynamics and formation of hybrid light-matter states (polaritons) while dynamically sweeping the metasurface resonance mode through PMMA's vibrational band. Figure 5a illustrates the hybrid polaritonic energy states, depicting lower polariton (LP) and upper polariton (UP), separated by the Rabi frequency ($\Omega$) associated with VSC at two distinct metasurface resonance levels. Figure 5b presents the simulated and measured spectral maps showing VSC-associated anti-crossing behavior as the metasurface temperature gradually increases and q-BIC-EIT mode traverses the PMMA band. The measured transmittance spectra across a 200 K temperature gradient in Figure 5c demonstrates that as the temperature increases and the q-BIC-EIT mode shifts to lower wavenumbers, the intensity of the UP gradually diminishes, while LP rises. We comprehensively characterized the polariton parameters in Figure 5d, where temperature dependent UP and LP peak wavenumber, transmittance amplitude, and linewidth data are presented. As the metasurface temperature increases and its resonance shifts to lower wavenumbers, LP couples more efficiently to the free space propagation than the UP, resulting in an enhanced LP transmission signal. Furthermore, we observed a systematic increase in polariton linewidths with temperature. This thermal broadening can be attributed to increased phonon interactions and scattering processes at elevated temperatures, contributing to reduction in coherence lifetime of the hybrid states.

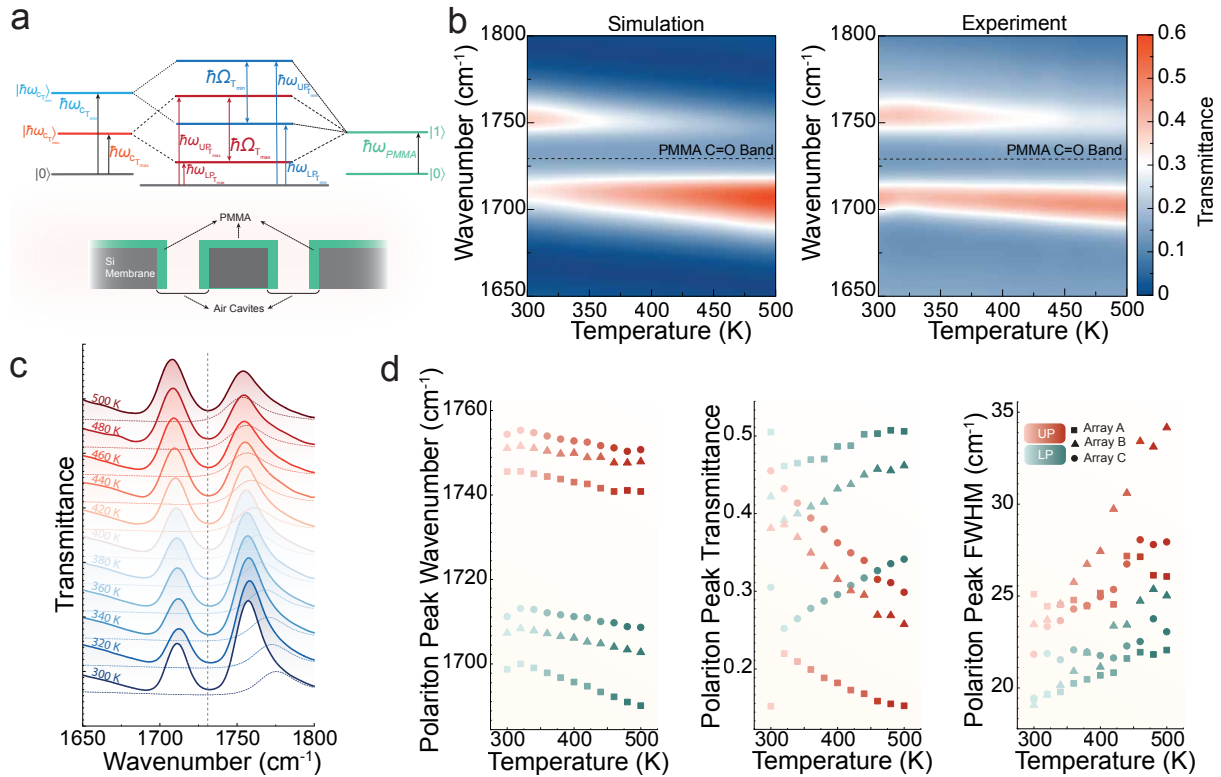

**Figure 5: Near-field vibrational strong coupling realized with dynamically tunable metasurfaces. a)** Upon coating the free-standing Si membrane metasurface with 50 nm thick PMMA, near-field light-matter interactions were studied. When the low-loss q-BIC-EIT mode spectrally overlaps with the PMMA's absorption band, vibrational strong coupling condition is satisfied, and hybrid light-matter states (polaritons) are generated. The schematic shows the upper (UP) and lower (LP) polariton states separated by the Rabi splitting ($\Omega$) at two different resonance frequencies (blue and red lines) of the same metasurface achieved by temperature tuning. **b)** Simulated and experimentally measured transmittance spectra of a PMMA coated metasurface at varying temperatures from 300 to 500 K. The temperature dependent transmittance maps show the Rabi splitting between the upper and lower polariton branches. The pronounced anti-crossing behavior captured by a single dynamically tunable metasurface confirms vibrational strong coupling over the entire temperature range. **c)** Measured FTIR spectra demonstrating the transmittance variation of LP and UP as the metasurface resonance is swept through the PMMA vibrational band. **d)** The upper (red) and lower (green) polariton peak wavenumber, maximum transmittance, and FWHM variations as a function of temperature measured from three different metasurfaces.

## Discussion

Our work demonstrates the first implementation of thermally tunable free-standing Si membrane metasurfaces exhibiting high-Q transmissive resonances in the mid-IR fingerprint region. By leveraging Si's thermo-optical properties, we achieved spatiotemporal control over metasurface resonance characteristics, demonstrating chemical analysis of remote objects without requiring bulky spectrometer equipment and strongly enhanced light-matter interactions leading to on-demand switching between polaritonic states.

Previous strategies to achieve broadband mid-IR tunability in high-Q dielectric metasurfaces often required mechanical components or spatially encoded designs. For instance, Ge metasurfaces were optically interrogated by sweeping the illumination angle to modulate the spectral positions of their q-BIC resonances. However, this approach depends on goniometric setups and moving parts, which limit their robust integration[13]. More recently, gradient metasurfaces have enabled broadband functionality by spatially encoding resonance frequencies across the chip via gradual variations in unit cell geometry[38,39]. However, these structures require large device footprints to cover broad spectral ranges and their continuous geometric scaling limits their resonance Q-factors. As a result, they are inherently less suited for real-time analyte detection, especially for analysis of spatially heterogeneous samples.

In contrast, our thermally tunable metasurface offers continuous spectral control at a fixed spatial location, eliminating the need for mechanical alignment or spatial scanning. This advantage is reflected in our experimental results: we observe a resonance redshift of 23.5 cm$^{-1}$ across a temperature range of 300–700 K, with a maximum tuning rate of 0.06 cm$^{-1}$/K. Importantly, the experimentally measured Q-factor reaches 186 ( at 350 K ), demonstrating that thermal tuning does not compromise the resonance sharpness. This enables high-resolution, non-contact detection of analytes, as shown by the clear identification of characteristic absorption bands of polystyrene (1450 and 1492 cm$^{-1}$) and PMMA (1730 cm$^{-1}$), even when the samples are positioned several millimeters above the metasurface. This spectrometer-free fingerprinting demonstration represents a significant advancement toward compact analytical instrumentation.

Moreover, our photonic device enables exceptional light-matter interaction capabilities in the strong coupling regime. Previously, vibrational strong coupling was studied in Fabry-Perot cavities, sweeping their resonances by angular tuning[40], or on low-loss dielectric metasurfaces by spatially varying resonances[14,15]. Our dynamically reconfigurable metasurfaces generate hybrid light-matter states, polaritons, where thermal and frequency variations act together to define the coupling dynamics and coherence properties. Such real-time characterization capabilities introduce new experimental means to further investigate the fundamental physics of hybrid light-matter states, without the need to sequentially measure spectra by varying spatial position or illumination angle.

In conclusion, we have demonstrated that our dynamically reconfigurable Si membrane metasurfaces provide a versatile platform for advanced mid-IR spectroscopy and sensing. By combining high-Q transmissive resonances, dynamic tunability and accessible photonic cavities make our approach particularly attractive for integration into compact, sensitive and field-deployable chemical analyzers. By addressing the key limitations of mechanically tuned and spatially distributed metasurfaces, we believe this work opens new opportunities in reconfigurable IR spectroscopy, vibrational polaritonics, and many other applications of actively tunable light–matter interaction systems.

## Methods

### Numerical Simulations

The finite element method (FEM) simulations were performed using COMSOL Multiphysics (COMSOL AB, Sweden) to calculate the transmittance spectrum, electric field enhancements, and displacement current density in Figures, 2, and 5. In all simulations, Floquet periodic boundary conditions were applied along the x- and y-axes, while the top and bottom boundaries in the z-direction were modeled as perfectly matched layers (PML). To excite the q-BIC-EIT resonances, an x-polarized plane wave was used at normal incidence. We used the temperature-dependent refractive index of silicon [35,41] to simulate the spectral response of the metasurface.

### Fabrication

The dynamically tunable membrane metasurfaces were fabricated using conventional semiconductor process steps. We first fabricated intrinsic single-crystal Si membranes measuring approximately 1 µm in thickness and 2.8 mm × 2.8 mm in area, supported by a 300 µm-thick Si frame with dimensions of 10 mm × 10 mm using photolithography and wet/dry Si etching processes. The metasurfaces were fabricated by patterning the free-standing Si membranes using electron beam lithography (EBL) for mask formation, followed by reactive ion etching (RIE) to anisotropically etch the Si and generate tilted rod voids with vertical, smooth sidewalls. Further details of the fabrication process can be found in Rosas et.al[15].

### Optical Characterization

Mid-infrared characterizations of the metasurfaces were performed in transmission mode using collimated, linearly polarized light at normal incidence on two different platforms: an FTIR system with a cryostat stage and a discrete frequency IR (DFIR) microscope with a ceramic electric heater. Figure 3 was generated using data collected by the FTIR spectrometer, coupled to an infrared microscope (Bruker Vertex 70 FTIR and Hyperion 2000). Transmittance spectra were acquired using linearly polarized light focused by a low-NA refractive objective (5×, 0.17 NA, Pike Technology, WI, USA). Collimated illumination was achieved by removing the bottom condenser, and detection was carried out using a liquid-nitrogen-cooled mercury-cadmium-telluride (MCT) detector. For temperature-dependent measurements, the metasurface was mounted directly on a cryostat stage (Linkam Scientific, FTIR600), and the temperature was swept from 300 K to 700 K.

For imaging-based spectral interrogation, we used a DFIR system integrated with four tunable quantum cascade lasers (QCL) on a microscope setup (Daylight Solutions, CA, USA), as shown in Figures 3, 4 and 5. The QCL microscope recorded the mid-IR spectrum in the range of 950 to 1800 cm$^{-1}$ with a spectral resolution of 2 cm$^{-1}$. Spectral data were collected using a 12.5× IR collection objective (0.7 NA) and detected with an uncooled microbolometer focal plane array (480 × 480 pixels), yielding a field of view of 650 μm × 650 μm. In this configuration, the metasurface was placed in direct contact with a ceramic heater controlled by a programmable power supply, allowing temperature variation from 300 K to 500 K. All measurements were referred to by air.

**Data extraction**

Peak resonance frequencies and Q-factors in Figures 2 and 3 were obtained by fitting the experimentally measured spectra with the following Fano fit from[42]

$$T = \left| ie^{i\phi} t_0 + \frac{\Gamma_R}{\Gamma_R + \Gamma_{NR} + i(\lambda - \lambda_{res})} \right|^2$$

Where $ie^{i\phi} t_0$ describes the background and the shape of the resonance. The Q-factor can then be calculated as:

$$Q = \frac{\lambda_{res}}{2(\Gamma_R + \Gamma_{NR})}$$

Where $\lambda_{res}$, $\Gamma_R$, and $\Gamma_{NR}$ are the resonance wavelength, radiative, and absorptive loss rates, respectively.

**Acknowledgment**
The authors gratefully acknowledge the use of facilities and instrumentation in the UW-Madison Wisconsin Center for Nanoscale Technology. The Center (wcnt.wisc.edu) is partially supported by the Wisconsin Materials Research Science and Engineering Center (NSF DMR-2309000) and the University of Wisconsin-Madison. Photonic metasurfaces were fabricated at the Center for Nanoscale Materials at the Argonne National Laboratory, a U.S. Department of Energy Office of Science User Facility, supported by the U.S. DOE, Office of Basic Energy Sciences, under Contract No. DE-AC02-06CH11357. F. K. acknowledges financial support from the Scientific and Technological Research Council of Turkiye (TÜBITAK) through the 2219 program under project number 1059B192300015. S. R. acknowledges financial support from the Wisconsin Distinguished Graduate Fellowship, made possible by The Grainger Foundation. M. A. K. acknowledges financial support from the Office of Naval Research (grant no.N00014-20-1-2297). Y. K. acknowledges financial support from the Australian Research Council (Grant No. DP210101292), International Technology Center Indo-Pacific (ITC IPAC) via Army Research Office (contract FA520923C0023). F. Y. acknowledges financial support from the U.S. National Science Foundation (Grant no. 2401616) and the U.S. National Institutes of Health (Grant no. R21EB034411)

**Data availability**
The data that support the findings of this study are available from the corresponding author upon reasonable request.

**References**

1 Gao X, Li X, Min W. Absolute Stimulated Raman Cross Sections of Molecules. *J Phys Chem Lett* 2023; **14**: 5701–5708.

2 Gao L, Qu Y, Wang L, Yu Z. Computational spectrometers enabled by nanophotonics and deep learning. *Nanophotonics* 2022; **11**: 2507–2529.

3 Wei J, Ren Z, Lee C. Metamaterial technologies for miniaturized infrared spectroscopy: Light sources, sensors, filters, detectors, and integration. *J Appl Phys* 2020; **128**: 240901.

4 Meng J, Weston L, Balendhran S, Wen D, Cadusch JJ, Unnithan RR *et al.* Compact Chemical Identifier Based on Plasmonic Metasurface Integrated with Microbolometer Array. *Laser Photonics Rev* 2022; **16**. doi:10.1002/lpor.202100436.

5 Wang A, Dan Y. Mid-infrared plasmonic multispectral filters. *Sci Rep* 2018; **8**: 11257.

6 Meng J, Balendhran S, Sabri Y, Bhargava SK, Crozier KB. Smart mid-infrared metasurface microspectrometer gas sensing system. *Microsyst Nanoeng* 2024; **10**: 74.

7 Tan X, Zhang H, Li J, Wan H, Guo Q, Zhu H *et al.* Non-dispersive infrared multi-gas sensing via nanoantenna integrated narrowband detectors. *Nat Commun* 2020; **11**: 5245.

8 Huang SH, Li J, Fan Z, Delgado R, Shvets G. Monitoring the effects of chemical stimuli on live cells with metasurface-enhanced infrared reflection spectroscopy. *Lab a Chip* 2021; **21**: 3991–4004.

9 Rosas S, Schoeller KA, Chang E, Mei H, Kats MA, Eliceiri KW *et al.* Metasurface-Enhanced Mid-Infrared Spectrochemical Imaging of Tissues. *Adv Mater* 2023; : e2301208.

10 Rodrigo D, Tittl A, Ait-Bouziad N, John-Herpin A, Limaj O, Kelly CV *et al.* Resolving Molecule-Specific Information in Dynamic Lipid Membrane Processes With Multi-Resonant Infrared Metasurfaces. 2018; **9**. doi:10.1038/s41467-018-04594-x.

11 Beisenova A, Adi W, Kang S, Germanson KB, Nam S, Rosas S *et al.* High-Precision Biochemical Sensing with Resonant Monocrystalline Plasmonic Ag Microcubes in the Mid-Infrared Spectrum. *ACS Nano* 2025; **19**: 13273–13286.

12 Tittl A, Leitis A, Liu M, Yesilkoy F, Choi D-Y, Neshev DN *et al.* Imaging-based molecular barcoding with pixelated dielectric metasurfaces. *Science* 2018; **360**: 1105–1109.

13 Leitis A, Tittl A, Liu M, Lee BH, Gu MB, Kivshar YS *et al.* Angle-multiplexed all-dielectric metasurfaces for broadband molecular fingerprint retrieval. *Sci Adv* 2019; **5**: eaaw2871.

14 Adi W, Rosas S, Beisenova A, Biswas SK, Mei H, Czaplewski DA *et al.* Trapping light in air with membrane metasurfaces for vibrational strong coupling. *Nat Commun* 2024; **15**: 10049.

15 Rosas S, Adi W, Beisenova A, Biswas SK, Kuruoglu F, Mei H *et al.* Enhanced biochemical sensing with high- Q transmission resonances in free-standing membrane metasurfaces. *Optica* 2025; **12**: 178.

16 Biswas SK, Adi W, Beisenova A, Rosas S, Arvelo ER, Yesilkoy F. From weak to strong coupling: quasi-BIC metasurfaces for mid-infrared light–matter interactions. *Nanophotonics* 2024; **13**: 2937–2949.

17 Shaltout AM, Shalaev VM, Brongersma ML. Spatiotemporal light control with active metasurfaces. *Science* 2019; **364**. doi:10.1126/science.aat3100.

18 Gu T, Kim HJ, Rivero-Baleine C, Hu J. Reconfigurable metasurfaces towards commercial success. *Nat Photonics* 2023; **17**: 48–58.

19 Jung C, Lee E, Rho J. The rise of electrically tunable metasurfaces. *Sci Adv* 2024; **10**: eado8964.

20 Zograf GP, Petrov MI, Makarov SV, Kivshar YS. All-dielectric thermonanophotonics. *Adv Opt Photonics* 2021; **13**: 643.

21 Chae HU, Shrewsbury B, Ahsan R, Povinelli ML, Kapadia R. GaAs Mid-IR Electrically Tunable Metasurfaces. *Nano Lett* 2024; **24**: 2581–2588.

22 Chung H, Hwang I, Yu J, Boehm G, Belkin MA, Lee J. Electrical Phase Modulation Based on Mid-Infrared Intersubband Polaritonic Metasurfaces. *Adv Sci* 2023; **10**: 2207520.

23 Rodrigo D, Limaj O, Janner D, Etezadi D, Abajo FJG de, Pruneri V *et al.* Mid-infrared plasmonic biosensing with graphene. *Science* 2015; **349**: 165–168.

24 Lee I-H, Yoo D, Avouris P, Low T, Oh S-H. Graphene acoustic plasmon resonator for ultrasensitive infrared spectroscopy. *Nat Nanotechnol* 2019; **14**: 313–319.

25 Wu C, Guo X, Duan Y, Lyu W, Hu H, Hu D *et al.* Ultrasensitive Mid-Infrared Biosensing in Aqueous Solutions with Graphene Plasmons. *Adv Mater* 2022; **34**: e2110525.

26 Hu H, Yang X, Zhai F, Hu D, Liu R, Liu K *et al.* Far-field nanoscale infrared spectroscopy of vibrational fingerprints of molecules with graphene plasmons. *Nat Commun* 2016; **7**: 12334.

27 Yao Y, Shankar R, Kats MA, Song Y, Kong J, Loncar M *et al.* Electrically Tunable Metasurface Perfect Absorbers for Ultrathin Mid-Infrared Optical Modulators. *Nano Lett* 2014; **14**: 6526–6532.

28 Hu Y, Bai Y, Zhang Q, Yang Y. Electrically controlled molecular fingerprint retrieval with van der Waals metasurface. *Appl Phys Lett* 2022; **121**: 141701.

29 Duan J, Alfaro-Mozaz FJ, Taboada-Gutiérrez J, Dolado I, Álvarez-Pérez G, Titova E *et al.* Active and Passive Tuning of Ultranarrow Resonances in Polaritonic Nanoantennas. *Adv Mater* 2022; **34**: e2104954.

30 Leitis A, Heßler A, Wahl S, Wuttig M, Taubner T, Tittl A *et al.* All-Dielectric Programmable Huygens' Metasurfaces. *Adv Funct Mater* 2020; **30**: 1910259.

31 Zhou C, Qu X, Xiao S, Fan M. Imaging Through a Fano-Resonant Dielectric Metasurface Governed by Quasi--bound States in the Continuum. *Phys Rev Appl* 2020; **14**: 044009.

32 Kim C, Kim Y, Lee M. Laser-Induced Tuning and Spatial Control of the Emissivity of Phase-Changing Ge2Sb2Te5 Emitter for Thermal Camouflage. *Adv Mater Technol* 2022; **7**. doi:10.1002/admt.202101349.

33 Conrads L, Heßler A, Völkel L, Wilden K, Strauch A, Pries J *et al.* Infrared Resonance Tuning of Nanoslit Antennas with Phase-Change Materials. *ACS Nano* 2023; **17**: 25721–25730.

34 Aigner A, Ligmajer F, Rovenská K, Holobrádek J, Idesová B, Maier SA *et al.* Engineering of Active and Passive Loss in High-Quality-Factor Vanadium Dioxide-Based BIC Metasurfaces. *Nano Lett* 2024; **24**: 10742–10749.

35 Li HH. Refractive index of silicon and germanium and its wavelength and temperature derivatives. *J Phys Chem Ref Data* 1980; **9**: 561–658.

36 Holdman GR, Jaffe GR, Feng D, Jang MS, Kats MA, Brar VW. Thermal Runaway of Silicon-Based Laser Sails. *Adv Opt Mater* 2022; **10**. doi:10.1002/adom.202102835.

37 Koshelev K, Lepeshov S, Liu M, Bogdanov A, Kivshar Y. Asymmetric Metasurfaces with High-Q Resonances Governed by Bound States in the Continuum. *Phys Rev Lett* 2018; **121**: 193903.

38 Richter FU, Sinev I, Zhou S, Leitis A, Oh S, Tseng ML *et al.* Gradient High-Q Dielectric Metasurfaces for Broadband Sensing and Control of Vibrational Light-Matter Coupling. *Adv Mater* 2024; **36**: e2314279.

39 Aigner A, Weber T, Wester A, Maier SA, Tittl A. Continuous spectral and coupling-strength encoding with dual-gradient metasurfaces. *Nat Nanotechnol* 2024; **19**: 1804–1812.

40 Xiang B, Ribeiro RF, Du M, Chen L, Yang Z, Wang J *et al.* Intermolecular vibrational energy transfer enabled by microcavity strong light–matter coupling. *Science* 2020; **368**: 665–667.

41 Harris TR. Optical properties of Si, Ge, GaAs, GaSb, InAs, and InP at elevated temperatures. 2010.

42 Fan S, Suh W, Joannopoulos JD. Temporal coupled-mode theory for the Fano resonance in optical resonators. *J Opt Soc Am A, Opt, image Sci, Vis* 2003; **20**: 569–72.